\documentclass[graybox]{svmult}

\usepackage{mathptmx}       
\usepackage{helvet}         
\usepackage{courier}        
\usepackage{type1cm}        
\usepackage{makeidx}         
\usepackage{graphicx}        
\usepackage{multicol}        
\usepackage[bottom]{footmisc}

\usepackage[backend=bibtex,
	    natbib=true,
	    style=numeric-comp,
	    url=false,
            isbn=false,
            doi=false,
            firstinits=true,
            sorting=none,
            hyperref=true]{biblatex}
\renewbibmacro{in:}{}

\usepackage{chemarr,placeins}
\DeclareMathAlphabet{\mathcal}{OMS}{zplm}{m}{n}

\usepackage[np]{numprint}
\npstyleenglish
\nprounddigits{2}
\usepackage{amsmath}
\usepackage{amssymb}
\usepackage{numprint}
\usepackage{todonotes}
\usepackage{subfig}
\usepackage{nicefrac}

\newcommand{\propensity}{\gamma}
\newcommand{\distr}{\Pi}
\newcommand{\distrCond}{\Pi_\infty}
\newcommand{\mol}{x}

\newcommand{\Ex}[1]{\ensuremath{\textstyle E\!\left(#1\right)}}
\newcommand{\rc}{\ensuremath{k}}
\newcommand{\nos}{\ensuremath{N_S}}
\newcommand{\tror}{{N}}
\newcommand{\nor}{\ensuremath{R}}
\newcommand{\rn}{\ensuremath{r}}
\newcommand{\maxentnom}{\ensuremath{K}}

\newcommand{\mcProcess}{\vec{X}}
\newcommand{\mcState}{\vec{x}}
\newcommand{\mcProcY}{\vec{Y}}
\newcommand{\mcProcZ}{\vec{Z}}
\newcommand{\mcStateY}{\vec{y}}
\newcommand{\mcStateZ}{\vec{z}}
\newcommand{\changeVector}{v}
\newcommand{\distrFamily}{\mathcal{G}}
\newcommand{\maxent}{q}

\makeindex             

\begin{document}

\title*{Distribution approximations for the chemical master equation: comparison of the method of moments and the system size expansion}
\titlerunning{Distribution approximations for the chemical master equation}  
\author{Alexander Andreychenko \and Luca Bortolussi \and Ramon Grima \and Philipp Thomas \and Verena Wolf}
\authorrunning{Andreychenko et al.}  
\institute{Alexander Andreychenko \at Modelling and Simulation Group, Saarland University, Germany, \email{makedon@cs.uni-saarland.de}
\and Luca Bortolussi \at Modelling and Simulation Group, Saarland University, Germany and Department of Mathematics and Geosciences, University of Trieste, Italy, \email{luca@dmi.units.it}
\and Ramon Grima \at School of Biological Sciences, University of Edinburgh, UK, \email{ramon.grima@ed.ac.uk}
\and Philipp Thomas \at School of Mathematics and School of Biological Sciences, University of Edinburgh, UK, \email{philipp.thomas@ed.ac.uk}
\and Verena Wolf \at Modelling and Simulation Group, Saarland University, Germany, \email{wolf@cs.uni-saarland.de}
}
%
%
\maketitle


\abstract{The stochastic nature of chemical reactions  involving randomly fluctuating population sizes has lead to a growing research interest in discrete-state stochastic models and their analysis. A widely-used  approach 
is the description of the temporal evolution of the system in terms of a chemical master equation (CME). 
In this paper we study two approaches for approximating the underlying probability distributions of the CME. The first approach is based on an integration of the statistical moments and the reconstruction of the distribution based on the maximum entropy principle.
The second approach relies on an analytical approximation of the probability distribution of the CME using the system size expansion, considering higher-order terms than the linear noise approximation. We consider gene expression networks with unimodal and multimodal protein distributions to compare the accuracy of the two approaches. We find that both methods provide accurate approximations to the distributions of the CME while having different benefits and limitations in applications.}

\section{Introduction}
\label{sec:intro}


It is widely recognised that noise plays a crucial role in shaping the behaviour of biological systems \cite{arkin97,swain2002,perkins09,elowitz02}. Part of such noise can be explained by intrinsic fluctuations of molecular concentrations inside a living cell, caused by the randomness of biochemical reactions, and fostered by the low numbers of certain molecular species \cite{swain2002}. 
As a consequence of this insight, stochastic modelling has rapidly become very popular \cite{wilkinson06}, dominated by Markov models based on the  Chemical Master Equation (CME) \cite{wilkinson06,gillespie77}. 

%

The CME represents a system of differential equations that specifies the time evolution of a discrete-state  stochastic model that explicitly accounts for the discreteness and randomness of molecular interactions. 
It has therefore been widely used to model gene regulatory networks, signalling cascades and other intracellular processes which are significantly affected by the stochasticity inherent in reactions involving low number of molecules~\cite{maheshri2007living}.

A solution of the CME yields the probability distribution over population vectors that count the
number of molecules of each chemical species. While a numerical solution of the CME is rather straight-forward, i.e., via a truncation of the state space~\cite{munsky2006,FAUIET},
for most networks the combinatorial complexity of the underlying state space renders efficient numerical integration infeasible. Therefore, the stochastic simulation algorithm (SSA), a Monte-Carlo technique, is commonly used to derive statistical estimates of the corresponding state probabilities \cite{gillespie2007}.

An alternative to stochastic simulation is to rely on approximation methods, that can provide fast and accurate estimates of some aspects of stochastic models. Typically, most approximation methods focus on the estimation of moments of the distributions\cite{elf2003,grima2010,thomas2013,Engblom,gillespie_cs_moment_closure,StumpfJournal}.  However, two promising approaches for the approximate computation of the distribution underlying the CME have recently been developed, whose complexity is independent of the molecular population sizes. 


The first method is based on the inverse problem, i.e., reconstructing the probability distribution from its moments. To this end, a closure on the moment equations is employed which yields an approximation of the evolution of all moments up to order $\maxentnom$ of the joint distribution \cite{Engblom,gillespie_cs_moment_closure,StumpfJournal}.
Thus, instead of solving one equation per population vector 
we solve $\sum_{k=1}^{\maxentnom} {\nos+k-1\choose k}$ 
equations if $\nos$ is the number of different chemical species.
Given the (approximate) moments at the final time instant, it is possible to reconstruct the corresponding
marginal probability distributions using the maximum entropy principle~\cite{andreychenko_mikeev_wolf_mc_based_reconstruction,alexander2014}.
The reconstruction requires the solution of a nonlinear constrained optimization problem. 
Nevertheless, the integration of the moment equations and the reconstruction of the underlying distribution
can for most systems be carried out very efficiently and thus allows a fast approximation of the CME. 
This is particularly useful if parameters of the reaction network have to be estimated based on observations
since most likelihoods can be determined if the marginal distributions are given.

The second method, which is based on van Kampen's system size expansion \cite{vanKampen}, does not resort to moments but instead represents a direct analytical approximation of the probability distribution of the CME. Unlike the method of moments, the technique assumes that the distribution can be expanded about its deterministic limit rate equations using a small parameter called the system size. For biochemical systems, the system size coincides with the volume to which the reactants are confined. The leading order term of this expansion is  given by the linear noise approximation which predicts that the fluctuations about the rate equation solution are approximately Gaussian distributed \cite{vanKampen} in agreement with the central limit theorem valid for large number of molecules. For low molecule numbers, the non-Gaussian corrections to this law can be investigated systematically using the higher order terms in the system size expansion. A solution to this expansion has recently been given in closed form as a series of the probability distributions in the inverse system size \cite{thomas2015}. Although in general the positivity of this distribution approximation cannot be guaranteed, it often provides simple and accurate analytical approximations to the non-Gaussian distributions underlying the CME.   


Since many reaction networks involve very small molecular populations, it is often questionable whether the system size expansion and moment based approaches can appropriately capture their underlying discreteness. For example, the state of a gene that can either be 'on' or 'off' while the number of mRNA molecules is of the order of only a few tens on average. In these situations, hybrid approaches are more appropriate, and supported by convergence results in a hybrid sense in the thermodynamic limit \cite{luca2015}. A hybrid moment approach for the solution of the CME integrates a small master equation for the small populations while a system of conditional moment equations is integrated for the large populations which is coupled with the equation for the small populations. If the equations of unlikely states of the small populations are ignored, a more efficient and accurate approximation of the CME can be obtained compared to the standard method of moments.
Similarly, a conditional system size expansion can be constructed that tracks the probabilities of the small populations and applies the system size expansion to the large populations conditionally. Presently, such an approach has employed the linear noise approximation for gene regulatory networks with slow promoters invoking timescale separation \cite{thomas2014}. The validity of a conditional system size expansion including higher than linear noise approximation terms is however still under question. 


Given these recent developments, it remains to be clarified how these two competing approximation methods perform in practice. Here, we carry out a comparative study between numerical results obtained using the methods of moments and analytical results obtained from the system size expansion for two common gene expression networks. The outline of the manuscript is the following: In Section \ref{sec:kinetics} we will briefly review the CME formulation. Section \ref{sec:mc} outlines the methods of (conditional) moments and the reconstruction of the probability distributions using the maximum entropy principle. In Section \ref{sec:sse} the approximate solution of the CME using the SSE is reviewed. We then carry out two detailed case studies in Section \ref{sec:results}. In the first case study, we investigate a model of a constitutively expressed gene leading to a unimodal protein distribution. In a second example we study the efficiency of the described hybrid approximations using the method of conditional moments and the conditional system size expansion for the prediction of multimodal protein distributions from the expression of a self-activating gene. These two example are typical scenarios encountered in more complex models, and as such provide ideal benchmarks for a qualitative and quantitative comparison of the two methods. We conclude with a discussion in Section \ref{sec:discussion}.

%
%


\section{Stochastic chemical kinetics}
\label{sec:kinetics}

A biochemical reaction network is specified by a set of $\nos$ different chemical species $S_1,\ldots,S_{\nos}$ and by a set of $\nor$ reactions of the form 
$$
\quad\quad   {\ell}_{{\rn},1}^{-} S_1+\ldots+  {\ell}^{-}_{{\rn},{\nos}} S_{\nos} \xrightarrow{\rc_j}  {\ell}_{{\rn},1}^{+} S_1+\ldots+  {\ell}_{{\rn},\nos}^{+} S_{\nos},
\quad 1\le {\rn} \le \nor.
$$
Given a reaction network, we define  a continuous-time Markov chain $\{\mcProcess(t),t\ge 0\}$, where
   $\mcProcess(t)=(X_1(t),\ldots,X_{\nos}(t))$ is a random vector   
whose $i$-th entry $X_i(t)$ is the number of molecules of type $S_i$.
If $\mcProcess(t)=(x_1,\ldots,x_{\nos})\in\mathbb N_0^{\nos}$ is the state of the process 
at time $t$ and $x_i\ge {\ell}_{{\rn},i}^{-}$ for all $i$, then the ${\rn}$-th 
reaction corresponds to a possible transition from state $\mcState$ to state $\mcState+\vec{\changeVector}_r$ 
where $\vec{\changeVector}_r$ is the change vector with entries 
$\changeVector_{{\rn},i} =  {\ell}_{{\rn},i}^{+}-{\ell}_{{\rn},i}^{-} \in \mathbb Z^{\nos}$.
The rate of the reaction is given by the propensity function $\propensity_r(\mcState)$, with $\propensity_r(\mcState)\mathrm{d}t$ being the probability of a reaction of index ${\rn}$ occurring in a time instant $\mathrm{d}t$, assuming that the reaction volume is well-stirred.
 The most common form of propensity function follows from the principle of mass action and depends on the volume $\Omega$ to which the reactants are confined, 
 $\propensity_r(\mcState):=\Omega \rc_r\prod_{i=1}^{\nos} \Omega^{-\ell_{{\rn},i}^{-}} {x_i\choose \ell_{{\rn},i}^{-}}$, 
 as it can be derived from physical kinetics~\cite{gillespie77,gillespie2009}. 	

%

We now fix the initial condition $\vec{x}_0$ of the Markov process to be deterministic and let 
$\distr (\mcState,t)=\mathrm{Prob}(\mcProcess(t)=\mcState \, | \, \mcProcess(0)=\mcState_0)$ for $t\ge 0$.
The time evolution of $\distr(\mcState,t)$ is governed by the Chemical Master Equation (CME) as
\begin{equation}
\label{eq:master}
	\frac{\mathrm{d} \distr \left( \mcState,t \right)}{\mathrm{d} t}
	= \sum_{{\rn}=1}^{\nor} \left(\propensity_r(\mcState-\vec{\changeVector}_r)\distr(\mcState-\vec{\changeVector}_r,t)
	 -\propensity_r(\mcState)\distr(\mcState,t)\right).
\end{equation}
We remark that the probabilities $\distr(\mcState,t)$ are uniquely determined because 
we consider the equations of all states that are reachable from the initial conditions.

\section{Method of moments}
\label{sec:mc}
For most realistic examples the number of reachable states is  extremely large or even
infinite, which renders an efficient numerical integration of Eq.~\eqref{eq:master}
impossible. 
An approximation of the moments of the distribution over time can be obtained
by considering the corresponding moment equations that describe the dynamics
of the first  $\maxentnom$ moments for some finite number $\maxentnom$.
We refer to this approach as the method of moments (MM).
In this section we will briefly discuss the derivation of the moment equations following
the lines of Ale et al.~\cite{StumpfJournal}.
%

  Let $\vec{f}:\mathbb N^{\nos}_0\to \mathbb R^{\nos}$ be a function that is independent of $t$.
 In the sequel we will exploit the following relationship, 
\begin{equation}\label{eq:expgen}
\begin{array}{lcl}
  \frac{d}{dt}\Ex{\vec{f}(\vec{X}(t))} &=& \sum\limits_{\vec{x}}
\vec{f}(\vec{x})\cdot  \frac{d}{dt} \distr(\vec{X}{(t)}=\vec{x})\\[2ex]
&=& \sum\limits_{{\rn}=1}^{\nor} \Ex{\propensity_{\rn}(\vec{X}(t)) \cdot \left( \vec{f}(\vec{X}(t) +  \vec{\changeVector}_{\rn}) - \vec{f}(\vec{X}(t)) \right)}.
\end{array}
\end{equation}
 For $\vec{f}(\vec{x})=\vec{x}$ this yields a system of equations for the population means
\begin{equation}\label{eq:mean1}
\begin{array}{lcl}
  \frac{d}{dt}\Ex{\vec{X}(t)} &=&   \sum\limits_{{\rn}=1}^{\nor} \vec{\changeVector}_{\rn} \Ex{\propensity_{\rn}(\vec{X}(t))}.
\end{array}
\end{equation}
Note that the system of ODEs in Eq.~\eqref{eq:mean1} is only closed
if at most monomolecular mass action reactions 
($\sum_{i=1}^{\nos} {\ell}^{-}_{{\rn},i}  \le 1$) are involved.
For most networks  the latter condition is not true and higher order moments appear
on the right side.
Let us write $\mu_i(t)$ for $\Ex{X_i(t)}$ and $\vec{\mu}(t)$
for the vector with entries $\mu_i(t)$, $1\le i\le {\nos}$.
Then a Taylor expansion of the function $\propensity_{\rn}(\vec{X}(t))$ about
the mean $\Ex{\vec{X}(t)}$ yields
\begin{equation}
 \begin{array}{rcl}
 \Ex{\propensity_{\rn}(\vec{X})}& =& \propensity_{\rn}(\vec{\mu}) + \frac{1}{1!}  \sum_{i=1}^{\nos}\Ex{X_i-\mu_i}\frac{\partial}{\partial \mu_i}\propensity_{\rn}(\vec{\mu})\\[1ex]
 &&+ \frac{1}{2!} \sum_{i=1}^{\nos} \sum_{k=1}^{\nos}\Ex{(X_i-\mu_i)(X_k-\mu_k)} \frac{\partial^2}{\partial \mu_i \partial \mu_k}\propensity_{\rn}(\vec{\mu})+\ldots,
 \end{array}
\end{equation}
 where we omitted $t$ in the equation to improve the readability.
 Note that $\Ex{X_i(t)-\mu_i}=0$ and assuming 
 that all propensities follow the law of mass action and all reactions
  are at most bimolecular, the terms of order three and more disappear. 
 By letting $C_{ik}$ be the covariance
 $\Ex{(X_i(t)-\mu_i)(X_k(t)-\mu_k)}$ we get
 \begin{equation}\label{eq:meanrate}
 \begin{array}{rcl}
 \Ex{\propensity_{\rn}(\vec{X})}& =& \propensity_{\rn}(\vec{\mu}) +  \frac{1}{2} \sum_{i=1}^{\nos} \sum_{k=1}^{\nos} C_{ik}\frac{\partial^2}{\partial \mu_i \partial \mu_k}\propensity_{\rn}(\vec{\mu}).
 \end{array}
 \end{equation}
 Next, we derive an equation for the covariances by first exploiting the relationship
\begin{equation}\label{eq:cov}
 \frac{d}{dt} C_{ik} =  \frac{d}{dt} \Ex{X_iX_k}- \frac{d}{dt}(\mu_i\mu_k) =
  \frac{d}{dt} \Ex{X_iX_k}- (\frac{d}{dt}\mu_i)\mu_k-\mu_i(\frac{d}{dt}\mu_k),
 \end{equation}
 and if we couple this equation with the equations for the means, the only unknown term
 that remains is the derivative $\frac{d}{dt} \Ex{X_iX_k}$ of the second moment.
  For this, we can apply the same strategy as before by using Eq.~\eqref{eq:expgen}
 for the test function
$f(\vec{x}):=\propensity_{\rn}(\vec{x})x_i$
 and consider the Taylor expansion about the mean. 
  Here, it is important to note that moments of order three come into play since derivatives
  of order three of $f(\vec{x}):=\propensity_{\rn}(\vec{x})x_i$ may be nonzero. 
 It is possible to take these terms into account
 by deriving additional equations for moments of order three and higher.
 These equations will then include moments of even higher order such that
 theoretically we end up with an infinite system of equations. 
 Different strategies to close the equations have been proposed in the literature~\cite{Whittle1957normalapprox,MatisCumulantTruncation2002,Krishnarajah2005novelapprox,singh2006lognormal,singh2011approximate}. 
 Here we consider a low dispersion closure  and assume that 
 all moments of order  $>\maxentnom$
 that are centered
 around the mean are equal to zero. 
 E.g. if we choose $\maxentnom=2$, then
 we can obtain the closed system of equations that does not include
 higher-order terms.
Then we can integrate the time evolution of the means and that of the covariances
 and variances.

In many situations, the approximation provided 
by the MM approach is very accurate even if only 
the means and covariances are considered. 
In general, however, numerical results show that the approximation tends to become worse 
if systems  exhibit complex behavior such as multistability or oscillations
and the resulting equations may become very stiff~\cite{Engblom}.
For some systems increasing the number of moments improves the accuracy~\cite{StumpfJournal}. Generally, however such a convergence is not seen \cite{Schnoer}, except in the limit of large volumes~\cite{grima2012study}.

\subsection{Equations for conditional moments}
For many reactions networks a hybrid moment approach, called 
method of conditional moments (MCM), can be more advantageous, in 
which we decompose  $\mcProcess(t)$ into 
small and large  populations.   
The reason is that small populations (often describing the activation state of a gene)
have distributions that assign a significant mass of probability
 only to a comparatively small
number of states. 
 In this case we can integrate the probabilities directly to get a more accurate
approximation of the CME compared to an integration of the moments.

Formally, we  consider $\mcProcess(t)=(\mcProcY(t),\mcProcZ(t))$,
where $\mcProcY(t)$ corresponds to the small, 
and $\mcProcZ(t)$ to the large populations. 
Similarly, we write $\mcState = (\mcStateY,\mcStateZ)$ 
for the states of the process
and $\vec{\changeVector}_{\rn}=(\hat{\vec{\changeVector}}_{\rn},\tilde{\vec{\changeVector}}_{\rn})$
for the change vectors,
${\rn} \in \{1,\ldots,{\nor}\}$.
Then, we condition on the state of the small populations and apply the MM to the conditional moments but not to the distribution of $\mcProcY(t)$ which we call the mode probabilities. 
Now, Eq.~\ref{eq:master} becomes 
\begin{equation}
\label{eq:master2}
\frac{d \distr (\mcStateY,\mcStateZ)}{dt} 
= \sum\limits_{{\rn}=1}^{\nor} ( \propensity_{\rn}(\mcStateY\!-\!\hat{\vec{\changeVector}}_{\rn},
\mcStateZ \!-\! \tilde{\vec{\changeVector}}_{\rn})  \distr(\mcStateY\!-\!\hat{
\vec{\changeVector}}_{\rn},\mcStateZ\!-\!\tilde{\vec{\changeVector}}_{\rn})
 -  \propensity_{\rn}(\mcStateY,\mcStateZ)\distr(\mcStateY,\mcStateZ)).
\end{equation}
 	Next, we sum over all possible $\mcStateZ$ to get the time evolution of the 
	marginal distribution $\hat \distr(\mcStateY)= \sum_{\mcStateZ} \distr(\mcStateY,\mcStateZ)$ 
	of the small populations.
	
	\begin{equation}\label{eq:smallcme}
	\begin{array}{l}
 \frac{d}{dt}\hat \distr(\mcStateY) =
 \sum\limits_{\mcStateZ} \sum\limits_{{\rn}=1}^{\nor} \propensity_{\rn}(\mcStateY-\hat{\vec{\changeVector}}_{\rn},\mcStateZ-\tilde{\vec{\changeVector}}_{\rn}) 
 \distr(\mcStateY-\hat{\vec{\changeVector}}_{\rn},\mcStateZ-\tilde{\vec{\changeVector}}_{\rn})
	- \sum\limits_{\mcStateZ} \sum\limits_{{\rn}=1}^{\nor} \propensity_{\rn}(\mcStateY,\mcStateZ) 
	\distr(\mcStateY,\mcStateZ)\\[2ex]
  \quad = \sum\limits_{{\rn}=1}^{\nor}  \hat \distr(\mcStateY-\hat{\vec{\changeVector}}_{\rn}) 
  E[\propensity_j(\mcStateY-\hat{\vec{\changeVector}}_{\rn},\mcProcZ)\mid \vec{Y}=\mcStateY-\hat{\vec{\changeVector}}_{\rn}]
	\sum\limits_{{\rn}=1}^{\nor}  \hat \distr(\mcStateY) 
	E[\propensity_{\rn}(\mcStateY,\mcProcZ)\mid \mcProcY=\mcStateY].
	\end{array}
	\end{equation}
	
	Note that in this small master equation that describes the change of the mode
	probabilities over time, the sum runs only over those reactions that modify 
	$\mcStateY$,
	since for all other reactions the terms cancel out. Moreover, on the right side
	we have only mode probabilities of neighboring modes and conditional
	expectations of the continuous part of the reaction rate. 
	For the latter, we can   use a Taylor expansion about the conditional
	population means. 
	Similar to Eq.~\eqref{eq:meanrate}, this yields an 
	equation that involves the conditional means and centered conditional
	moments of second order (variances and covariances).
	Thus, in order to close the system of equations, we need to derive equations
	for the time evolution of the  conditional means and centered conditional
	moments of higher order. 
	Since the mode probabilities may 
	become zero, we first derive an equation for the evolution of the partial means
	(conditional means multiplied by the probability of the condition)

		\begin{equation}
		\begin{array}{l}
			\frac{d}{dt}  \left(E[\mcProcZ \mid \mcStateY]\ \distr(\mcStateY) \right)
			= \sum\limits_{\mcStateZ} \mcStateZ \frac{d}{dt}  \distr(\mcStateY,\mcStateZ)\\[1ex]
		\hspace{2ex}	 = 
		\sum\limits_{{\rn}=1}^{\nor}   
 			 E[(\mcProcZ + \tilde{\vec{\changeVector}}_{\rn}) 
			 	\propensity_{\rn}(\mcStateY-\hat{\vec{\changeVector}}_{\rn}, \mcProcZ)
			 	\mid \mcStateY-\hat{\vec{\changeVector}}_{\rn}]\ 
			 	\distr(\mcStateY-\tilde{\vec{\changeVector}}_{\rn})\\[1ex]
			\hspace{2ex} - 
			\sum\limits_{{\rn}=1}^{\nor}   
			E[\mcProcZ \propensity_{\rn}(\mcStateY,\mcProcZ)	\mid \mcStateY]\ \distr(\mcStateY)
			,
		\end{array}
		\end{equation}
	where in the second line we applied Eq.~\eqref{eq:master2} and simplified
	the result.
	The conditional expectations 
	$E[(\mcProcZ + \tilde{\vec{\changeVector}}_{\rn}) 
				 	\propensity_{\rn}(\mcStateY-\hat{\vec{\changeVector}}_{\rn}, \mcProcZ)
				 	\mid \mcStateY-\hat{\vec{\changeVector}}_{\rn}]$
		and 
		$E[\mcProcZ \propensity_{\rn}(\mcStateY,\mcProcZ)	\mid \mcStateY]$
	are then replaced by their Taylor expansion about the conditional 
	means such that the equation involves only
	conditional means and  higher centered conditional moments~\cite{MCM_Hasenauer_Wolf}.
	For higher centered conditional moments, similar equations can be derived. If all
	centered conditional moments of order higher than $\maxentnom$ are assumed to be zero,
	the result is a (closed) system of differential algebraic equations (algebraic equations are
	obtained whenever a mode probability 
	$\distr(\mcStateY)$ is equal to zero).
	However, it is possible to transform the system of differential algebraic equations 
	into a system of (ordinary) differential   equations
	after truncating modes with insignificant probabilities.
	Then we can get an accurate 
	approximation of the solution 
	after applying standard numerical integration methods.
	For the case study in Section~\ref{sec:results} we constructed the ODE system 
	using the tool  
	SHAVE~\cite{hscc11}
	which implements 
	a truncation based approach
	and solves it using explicit Runge-Kutta method.

\subsection{Maximum entropy distribution reconstruction}
\label{s:maxent}

Given the first $\maxentnom+1$ (approximated) moments of a distribution 
$\Ex{X_{P}^0}, \Ex{X_{P}^1}, \allowbreak \ldots, \allowbreak \Ex{X_{P}^K}$
it is possible  
to reconstruct the corresponding probability distribution. 
Since a finite number of moments defines a set of distributions,
we apply  the \emph{maximum entropy principle}
where we choose  among  the distributions that
fulfill the moment equations, the one that maximizes the entropy. 
For instance, the normal distribution is chosen among all continuous distributions with equal
mean and variance. 

In the sequel we describe how 
to obtain the one-dimensional marginal probability distributions
of a reaction network
when we use the moments up to order $\maxentnom$.
We mostly follow Andreychenko et al.~\cite{andreychenko_mikeev_wolf_mc_based_reconstruction}
and simply
write $\mcProcess$ (and $\mcState$) for any random vector (and state) 
of at most $\nos$ molecular populations
at some fixed time instant $t$.
Given a sequence of 
$\maxentnom+1$ 
non-central moments\footnote{Non-central moments can be easily obtained from central ones by 
simple algebraic transformations. For instance, the second non-central moment is obtained from variance and mean as $\mu^{(2)}_i = C_{ii} + \mu_i^2$.}
$\Ex{X^k}=\mu^{(k)}, k=0,1,\ldots,\maxentnom,$
the set $\distrFamily$ of allowed (discrete) probability distributions 
consists of all non-negative functions $g$ for which the following conditions hold:
\begin{equation}
\label{eq:momentconstr}
	\textstyle\sum\limits_x x^k g(x)   = \mu^{(k)}, k=0,1,\ldots,\maxentnom.
\end{equation}
Here $x$ ranges over possible arguments (usually $x \in \bbbn_0$) with positive probability.
Note that we have included the constraint $\mu_0=1$ in order to 
guarantee that $g$ is a probability distribution.	
According to the maximum entropy principle, we choose		 
the distribution  $q \in \distrFamily$ that 
maximizes the entropy $H(g)$, i.e.
\begin{equation}
\label{eq:maxShannonProblem}
\begin{array}{c}
	\maxent \!=\! \arg \max\limits_{\maxent \in \distrFamily} H(g)
			\!=\! \arg \max\limits_{g\in \distrFamily} \left( -\sum\limits_x g(x) \ln{g(x)}\right)
			.
\end{array}
\end{equation}
The problem of finding the   maximum entropy distribution
is a nonlinear constrained optimization problem that
can be addressed by considering the Lagrangian	functional
$$
\begin{array}{c}
	\mathcal{L}(g,\lambda)
	= H(g) - \sum\limits_{k=0}^{\maxentnom} \lambda_k
			\left( \sum\limits_x x^k g(x)  - \mu^{(k)} \right),
\end{array}
$$
where $\lambda=(\lambda_0,\ldots,\lambda_\maxentnom)$ are the corresponding Lagrangian multipliers. The 
maximum of the unconstrained Lagrangian $\mathcal{L}$
corresponds to the solution of the constrained maximum entropy 
problem~\eqref{eq:maxShannonProblem}.
Note that setting the derivatives
of $\mathcal{L}(g,\lambda)$ w.r.t.\ $\lambda_k$, to zero results in the 
moment constraints.
	The   general form of the maximum   is obtained by  setting 
	$\frac{\partial \mathcal{L}}{\partial g(x)}$ to zero which yields
	$$
 		\textstyle	\maxent(x) = \textstyle	  \exp \left( -1 -\sum\limits_{k=0}^{\maxentnom} \lambda_k x^k \right) 
			=\frac{1}{Z} \exp \left( -\sum\limits_{k=1}^{\maxentnom} \lambda_k x^k \right),
 	$$
where
\begin{equation}
\label{eq:normalizationConst}
\textstyle Z = e^{1+\lambda_0}
      = \sum\limits_x 
      \exp \left( -\sum\limits_{k=1}^{M} \lambda_k x^k \right)
\end{equation}
is a normalization constant. The last equality in Eq.~\eqref{eq:normalizationConst} follows from the fact that $\maxent$ is a distribution and thus $\lambda_0$ is uniquely determined by
$\lambda_1,\ldots,\lambda_\maxentnom$.
Next we insert the above general form   into
the Lagrangian, thus transforming the problem into an
unconstrained convex minimization problem of the dual function w.r.t
the  variables $\lambda_k$. This yields the dual function
\begin{equation}
\label{eq:dualFunction}
\textstyle\Psi(\lambda)=\ln Z + \sum\limits_{k=1}^{\maxentnom} \lambda_k \mu^{(k)}.
\end{equation}
According to the Kuhn-Tucker theorem,
the solution
$\lambda^* = \arg \min \Psi(\lambda)$ of the minimization problem
determines the solution $\maxent$ 
of the original constrained optimization problem in Eq.~\eqref{eq:maxShannonProblem} (see~\cite{berger_maximum_1996}).
	We solve this unconstrained optimization problem
	using the Newton method from the MATLAB's
	numerical minimization package
	\texttt{minFunc},
	where we choose
	$\lambda^{(0)} = \left( 0, \ldots, 0 \right)$
	as an initial starting point 
	and use the approximated gradient and Hessian
	matrix.
	Since for the systems that we consider,
	the dual function 
	is convex~\cite{abramov2010multidimensional,wu2001fast,mead1984maximum},
	there exists a unique minimum 
	$\lambda^* = \left( \lambda^*_1, \ldots, \lambda^*_\maxentnom \right)$	
	where all first derivatives are
	zero and where the Hessian is positive definite. 
	The final results $ \lambda^*$ 
	of the iteration yields the distribution 
	\begin{center}
	\label{eq:maxent_reconstruction}
	$\tilde{\maxent}(x) = \exp ( -1 -\sum\limits_{k=0}^{\maxentnom} \lambda^*_k x^k)$,
	\end{center} 
	which is an approximation 
	of the marginal distribution 
  of the process at time $t$.
  
  The sequence of moments $\mu^{(k)}$, $k=0,\ldots,K$
  obtained using MM or MCM serves
  as an input to the maximum entropy reconstruction procedure.
  Due to the high sensitivity with respect to the accuracy 
  of the highest order moment $\Ex{X_{P}^K}$,
  we compute all moments up to $\Ex{X_{P}^{K+1}}$ to get the
  better estimation but use moments only up to $\Ex{X_{P}^K}$
  in the entropy maximization.

	The maximum entropy approach provides a natural extension 
	of the moment-based integration methods of the CME. 
	It does not make any additional assumptions about the probability distribution 
	apart from the moment constraints and the corresponding optimization
	problem can be solved efficiently for one- and two-dimensional
	distributions. 
	The reconstruction of the distributions
	of higher dimension is more involved due to 
	numerical instabilities arising when using
	the ill-conditioned Hessian matrix~\cite{alldredge_adaptive_2014,abramov2010multidimensional}
	in the optimization procedure.
	As mentioned in the numerical results that we present in the sequel,
	problems may arise if the support of the distribution 
	(which serves as an input argument to the optimization procedure)
	is not chosen	adequately.
	The true support of the distribution often infinite
	and the reasonable truncation has to be used~\cite{tari2005unified}.
	The possible solution is addressed in~\cite{alexander2014} where we
	introduce an iterative heuristic-based procedure of the support approximation.
%
%
	However,
	generally this approach gives very accurate results relative to
	 the information about the distribution given by the moment
	constrains.


\section{System size expansion of the probability distribution}
\label{sec:sse}
We here describe the use of the system size expansion \cite{vanKampen} to obtain approximate but simple expressions for the probability distributions the CME. For simplicity, we will focus on the case of a single species and follow the approach by \emph{Thomas and Grima} \cite{thomas2015}. The system size expansion makes use of the macroscopic limit of the CME which is attained for large reaction volumes. Since large volumes imply large number of molecules when concentrations are held constant, the macroscopic concentration $[X]$ is described by the deterministic rate equation 
\begin{align}
 \label{eqn:REs}
 \frac{\mathrm{d} [X]}{\mathrm{d} t}=\sum_{r=1}^R \nu_r f_r^{(0)}([X]).
\end{align}
Note that here $\nu_r=\nu_{r,1}$ because we only consider a single species. A prerequisite for Eq. (\ref{eqn:REs}) to be the deterministic limit of the CME is that the rate functions satisfy the scaling 
\begin{align}
\label{eqn:propensityexpansion}
\propensity_j(\Omega [X],\Omega)=\Omega \bigl[ f^{(0)}_r([X]) + \Omega^{-1} f^{(1)}_r([X]) + \ldots \bigr],
\end{align}
where the first term in this series, {$f_r^{(0)}([X])= \lim_{\Omega\to\infty} \frac{\propensity_r(\Omega [X])}{\Omega}$,}
is the macroscopic rate function. Note that for an unimolecular reaction only the first term in Eq. (\ref{eqn:propensityexpansion}) is present while for a bimolecular one the first two terms are non-zero \cite{thomas2015}.

The expansion allows to characterize the deviations from this deterministic behaviour by successively taking into account higher order terms. Specifically, van Kampen proposed separating the dynamics of the molecular concentration into the deterministic part $[X]$ and a fluctuating component $\epsilon$ that reads
\begin{align}
\label{eqn:ansatz}
\frac{\mol}{\Omega}=[X] + \Omega^{-1/2}\epsilon.
\end{align}
This ansatz can be used to expand the CME in powers of the inverse square root of the volume by changing to $\epsilon$-variables. Assuming a continuous approximation, i.e., $\distr(\epsilon,t)=\distr(\Omega[X]+\Omega^{1/2}\epsilon,t)\Omega^{1/2}$, the CME becomes
\begin{align}
 \label{eqn:CMEexp}
 &\frac{\partial}{\partial t}  \distr(\epsilon,t) = \mathcal{L}_0 \distr(\epsilon,t) + \sum_{k=1}^\tror \Omega^{-k/2}\mathcal{L}_k  \distr(\epsilon,t) + O(\Omega^{-(\tror+1)/2}),
\end{align}
which essentially is a partial differential approximation when truncated after the $\tror^{th}$ term. It is shown in Ref. \cite{thomas2015} that the differential operators $\mathcal{L}_k$ can be obtained explicitly and are given by
\begin{align}
 \mathcal{L}_k = \sum_{s=0}^{\lceil k/2\rceil} \sum_{p=1}^{k-2(s-1)} \frac{\mathcal{D}_{p,s}^{k-p-2(s-1)}}{p!(k-p-2(s-1))!} (-\partial_\epsilon)^p \epsilon^{k-p-2(s-1)},
\end{align}
where the coefficients
\begin{align}
 \label{eqn:SSEcoeffs}
 \mathcal{D}_{p,s}^{q}= \sum_{r=1}^R (\nu_r)^p \frac{\partial^q f_r^{(s)}([X])}{\partial{[X]}^q},
\end{align}
depend only on the solution of the rate equation. We now will solve Eq. (\ref{eqn:CMEexp}) perturbatively by expanding the probability density as
\begin{align}
 \label{eqn:asymptoticExpansion}
 \distr(\epsilon,t)=\sum_{j=0}^\tror \Omega^{-j/2} \pi_j(\epsilon,t) + O(\Omega^{-(\tror+1)/2}).
\end{align}
%
%
Using the above series in Eq. (\ref{eqn:CMEexp}) and equating order $\Omega^0$ terms we find  
\begin{subequations}
\begin{align}
\label{eqn:LNA}
\left(\frac{\partial}{\partial t} - \mathcal{L}_0  \right) \pi_{0} = 0,
\end{align}
while equating terms to order $\Omega^{-j/2}$, we find
\begin{align}
  \label{eqn:systemofequations}
  \biggl(\frac{\partial}{\partial t}-\mathcal{L}_0\biggr) \pi_{j}(\epsilon) 
  = \mathcal{L}_1 \pi_{j-1} + \ldots + \mathcal{L}_j \pi_0,
\end{align}
\end{subequations}
for $j>0$.
The first equation, Eq. (\ref{eqn:LNA}), is called the linear noise approximation \cite{vanKampen} and its solution is a Gaussian distribution
\begin{align}
 \label{eqn:pi0}
 \pi_0(\epsilon,t)=\frac{1}{\sqrt{2\pi\sigma^2(t)}} \exp\left({-\frac{\epsilon^2}{2\sigma^2(t)}}\right), 
\end{align}
with zero mean meaning that the rate equation are valid on average. Its variance $\sigma^2(t)$ follows the equation
\begin{align}
 \label{eqn:lyapunov}
 \frac{\partial \sigma^2}{\partial t}  = 2 \mathcal{J}(t) \sigma^2 + \mathcal{D}_2^0(t),
\end{align}
where we have denoted by $\mathcal{J}(t)=\mathcal{D}_1^1(t)$ the Jacobian of the rate equation (\ref{eqn:REs}).

The system of partial differential equations (\ref{eqn:systemofequations}) can be obtained using the eigenfunctions of $\mathcal{L}_0$. In particular we can write $\pi_j(\epsilon,t)=\sum_{m=1}^{3j} a_m^{(j)}(t) \psi_m({\epsilon},t)\pi_0(\epsilon,t)$ where   $\psi_m(\epsilon,t)=\pi_0^{-1}(-\partial_\epsilon)^m \pi_0 = \frac{1}{\sigma^m} H_m\left(\frac{\epsilon}{\sigma}\right)$ and $H_m$ are the Hermite polynomials. The solution of the system size expansion is therefore
\begin{subequations}
\label{eqn:SSEbareSol}
\begin{align}
 \label{eqn:HermiteExpansion}
 \distr(\epsilon,t) =& \pi_0(\epsilon,t)\biggl( 1+\sum_{j=1}^{\tror} \Omega^{-j/2}\sum_{m=1}^{3j} a_m^{(j)}(t) \psi_m\left({\epsilon},t\right) \biggr) 
 + O(\Omega^{-(\tror+1)/2}).
\end{align}
Mathematically speaking the above equation represents an asymptotic series solution to the CME.
Equations for the coefficients can be obtained using the orthogonality of the Hermite polynomials, and are given the ordinary differential equations
\begin{align}
 \label{eqn:coefficientEquations}
 &\left( \frac{\partial}{\partial t}- n \mathcal{J} \right) a_n^{(j)} =\sum_{k=1}^j \sum_{m=0}^{3({j-k})} a_m^{(j-k)} \sum_{s=0}^{\lceil k/2\rceil} \sum_{p=1}^{k-2(s-1)} \mathcal{D}_{p,s}^{k-p-2(s-1)} \mathcal{I}_{mn}^{p,k-p-2(s-1)},
\end{align}
with
\begin{align}
 \label{eqn:integral}
 \mathcal{I}^{\alpha\beta}_{mn}
   =& \frac{\sigma^{\beta-\alpha+n-m}}{\alpha!}\sum_{s=0}^{\min({n-\alpha,m})}\binom ms  \frac{(\beta+\alpha+2s-(m+n)-1)!!}{(\beta+\alpha+2s-(m+n))!(n-\alpha-s)!},
\end{align}
\end{subequations}
and zero for odd $(\alpha+\beta)-(m+n)$. Note that $a_n^{(j)}=0$ when $n+j$ is odd. Explicit expressions for the probability density can now be evaluated to any desired order. Particularly simple and explicit solutions can be obtained in steady state by letting the time-derivative on the left hand side of Eq. (\ref{eqn:coefficientEquations}) go to zero. It follows that the first term in Eq. (\ref{eqn:HermiteExpansion}), the linear noise approximation ${\pi}_0$, describes the distribution in the infinite volume limit while for finite volumes, implying low number of molecules, the remaining terms have to be taken into account.

It is however the case that this approximation can become inaccurate for processes whose mean behaviour differs significantly from the rate equation. This is because van Kampen's ansatz, Eq. (\ref{eqn:ansatz}), uses $\epsilon$ to denote the fluctuations about that the average given by the solution of the rate equation $[X]$. For biochemical involving bimolecular reactions the propensities depend nonlinearly on the concentrations and hence the rate equations are only approximations to the true averages predicted by the CME. 
In applications it is important to account for these deviations from the rate equation solution and the linear noise approximation. We therefore calculate the true concentration mean and variance using the system size expansion a priori and then perform the expansion about the true mean. A posteriori, this leads to an expansion about the mean
\begin{align}
 \label{eqn:ansatzRen}
 \frac{\mol}{\Omega}=\underbrace{\biggl \langle \frac{\mol}{\Omega}\biggr\rangle}_{\text{mean}} + \underbrace{\biggl.\Omega^{-1/2}\bar{\epsilon}\biggr.}_{\text{fluctuations}},
\end{align}
which is different than the one proposed by van Kampen, Eq. (\ref{eqn:ansatz}), who expands the CME about the solution of the rate equation.
Here, the averages are calculated from $\bigl \langle \frac{\mol}{\Omega}\bigr\rangle = [X]+\Omega^{-1/2}\langle\epsilon\rangle$ such that $\bar{\epsilon}=\epsilon-\langle\epsilon\rangle$ is a centered random variable quantifying the fluctuations about the true average which can be obtained using the system size expansion. The result is an expansion about the mean (instead of the rate equation) which is given by 
\begin{subequations}
\label{eqn:RenSol}
\begin{align}
 \label{eqn:RenormalizedExpansion}
 \bar{\pi}(\bar{\epsilon},t) &= \bar{\pi}_0(\bar{\epsilon},t) + \sum_{j=1}^{\tror} \Omega^{-j/2}\sum_{m=1}^{3j} \bar{a}_m^{(j)} \psi(\bar{\epsilon},t)  \bar{\pi}_0(\bar{\epsilon},t) + O(\Omega^{-(\tror+1)/2}),
\end{align}
where $\bar{\pi}_0(\bar{\epsilon})$ is a Gaussian different from the linear noise approximation which is centered about the true mean instead of the rate equation. The coefficients in the above equation can be calculated from
\begin{align}
 \label{eqn:renormCoeffs}
 \bar{a}_m^{(j)}=\sum_{k=0}^{j} \, \sum_{n=0}^{3k} a_{n}^{(k)}\kappa_{m-n}^{(j-k)},
\end{align}
and 
\begin{align}
 \label{eqn:kappa}
 \kappa_j^{(n)} = 
 \frac{1}{n!}\sum_{m=0}^{\lfloor j/2 \rfloor} (-1)^{(j+m)} \sum_{k=j-2m}^{n-m} { n \choose k} 
   B_{k,j-2m}\left(\left\{\zeta!a_1^{(\zeta)}\right\}\right) B_{n-k,m}\left(\left\{\frac{\zeta!}{2}\bar{\sigma}_{(\zeta)}^2\right\}\right).
\end{align}
\end{subequations}
Here $a_1^{(j)}$ and $\bar{\sigma}_{(j)}^2= 2\bigl[a_2^{(j)}- {B_{j,2}(\{\zeta!a_1^{(\zeta)}\})}/{j!}\bigr]$ denote the coefficients in the expansions of mean and variance
\begin{subequations}
\label{eqn:sseMeanVar}
\begin{align}
\langle\epsilon\rangle &= \sum_{j=1}^\tror \Omega^{-j/2} a_1^{(j)}+ O(\Omega^{-(\tror+1)/2}),\\
{\bar{\sigma}^2}
  &= \sigma^2 + \sum_{j=1}^\tror \Omega^{-j/2} {\sigma^2_{(j)}}+ O(\Omega^{-(\tror+1)/2}),
\end{align}
\end{subequations}
respectively, and $B_{k,n}(\{y_\zeta\})$ denotes the partial Bell polynomials \cite{comtet1974} defined as
\begin{align}
  B_{n,k}(\{y_\zeta\}_{\zeta=1}^{n-k+1})= {\sum}' {n!\primfrac {over}j_1! \ldots j_{n-k+1}!} \left ({y_1 \primfrac
  {over}1!}\right )^{j_1}\ldots\left ({y_{n-k+1} \primfrac
  {over}(n-k+1)!}\right )^{j_{n-k+1}}.
\end{align}
Note that $\sum'$ denotes the summation over all sequences $j_1,\ldots,j_{n-k+1}$ of non-negative integers such that $j_1 + \protect \ldots + j_{n-k+1}=k$ and $j_1 + 2j_2 + \protect \ldots + ({n-k+1})j_{n-k+1}=n$. Note that the expansion about the mean has generally less non-zero coefficients because $\bar{a}_1^{(j)}=\bar{a}_2^{(j)}=0$. Note also that for systems with propensities that depend at most linearly on the concentrations, mean and variance are exact to order $\Omega^0$ (linear noise approximation), and hence for this case expansion (\ref{eqn:HermiteExpansion}) coincides with Eq. (\ref{eqn:RenormalizedExpansion}). In Section~\ref{sec:results} we show that typically a few terms in this expansion (\ref{eqn:RenormalizedExpansion}) are sufficient to capture the underlying distributions of the CME.

A particular relevant case namely the stationary solution of the CME turns out to be obtained quite straight-forwardly. For example, truncating after $\Omega^{-1}$-terms, from Eq. (\ref{eqn:sseMeanVar}) it follows that 
$\langle\epsilon\rangle = \Omega^{-1/2} a_1^{(1)}+O(\Omega^{-3/2})$ and $\bar{\sigma}^2=\sigma^2 + \Omega^{-1} (2 a_2^{(2)}-(a_1^{(1)})^2) + O(\Omega^{-3/2})$. Letting now the left hand side of Eq. (\ref{eqn:coefficientEquations}) go to zero, solving for the coefficients $a_n^{(j)}$ and using the solution in Eq. (\ref{eqn:renormCoeffs}) one finds that to order $\Omega^{-1/2}$ the only non-zero coefficient is
\begin{subequations}
\label{eqn:renormalizedCoeffSol}
\begin{align}
\bar{a}_3^{(1)} =& -\frac{\sigma^4 \mathcal{D}_1^{2}}{6\mathcal{J}}+\frac{\sigma^2 \mathcal{D}_2^{1}}{6\mathcal{J}}+\frac{\mathcal{D}_3^{0}}{18 \mathcal{J}},
\end{align}
while to order $\Omega^{-1}$, one finds
\begin{align}
\bar{a}_4^{(2)} =&
-\frac{\mathcal{D}_{4}^0}{96 \mathcal{J}}-\frac{\sigma ^2 \mathcal{D}_{3}^1}{24 \mathcal{J}}-\frac{\sigma ^4 \mathcal{D}_{2}^2}{16 \mathcal{J}}-\frac{\sigma^6 \mathcal{D}_{1}^3}{24 \mathcal{J}} -\bar{a}_{3}^{(1)} \left(\frac{3 \mathcal{D}_{2}^1}{8 \mathcal{J}}+\frac{3 \sigma ^2 \mathcal{D}_{1}^2}{4 \mathcal{J}}\right),\notag\\
\bar{a}_6^{(2)} =& \frac{1}{2} (\bar{a}_3^{\text{(1)}})^2.
\end{align}
\end{subequations}
The above equations determine the series solution of the CME, Eq. (\ref{eqn:RenormalizedExpansion}), in stationary conditions to order $\Omega^{-1}$.

\section{Results}
\label{sec:results}
In this section we will compare the (hybrid) method of moments (MM and MCM approach) and the (conditional) system size expansion (SSE approach) by considering the quality of the resulting distribution approximation.
In the former case we use the maximum entropy approach outlined in Section \ref{s:maxent} and the approximate solution obtained from the SSE in the latter case. We base our comparison on two simple but challenging examples. The first one describes the bursty expression of a protein which degrades via an enzyme-catalyzed reaction. The second example describes the expression of a protein activating its own expression resulting in a multimodal protein distribution.


\subsection{Example 1: Bursty protein production}
\label{sec:bursts}

In order to compare the performance of the two methods, we will employ a simple model of gene expression with enzymatic degradation described in Ref. \cite{thomas2012}.
The system itself is described by the following set of biochemical reactions:
\begin{align}
\label{reactions:bursts}
&\varnothing \xrightarrow{\rc_0} M \xrightarrow{\rc_1} \varnothing, \ \ M \xrightarrow{\rc_2} M+P, \notag\\
&P+E  \xrightleftharpoons[\rc_4]{\rc_3} C  \xrightarrow{\rc_5} E .
\end{align}
 Active protein degradation has been recently been recognized to be an important factor in skewing nuclear protein distributions in mammalian cells towards high molecule numbers \cite{giampieri2015}. 
In our model we explicitly account for mRNA which is important even when the mRNA half-life is shorter than the one of the corresponding protein \cite{shahrezaei2008}. Assuming binding and unbinding of $P$ and $E$ are fast compared to protein degradation, the degradation kinetics can be simplified as in Ref. \cite{giampieri2015,thomas2011}, resulting in a Michaelis-Menten like kinetic rate:
\begin{align}
 P \xrightarrow{f(\mol_P)} \varnothing,
\end{align}
with $f(\mol_P) =  \frac{\Omega\, v_M \mol_P}{\Omega K_M+\mol_P}$, where $\mol_P$ is the number of proteins and $K_M=\frac{\rc_4+\rc_5}{\rc_3}$ is the Michaelis-Menten constant. Less obviously for this reduction to hold one also requires $\rc_4 \gg \rc_5$ as shown in Refs. \cite{thomas2011,sanft2011}. Here rates are set to 
$k_0 = 8$, $k_1 =10$, $k_2 = 100$, $v_M = 100$, $K_M = 20$, $\Omega=1$. 
In particular, the reaction rates involving mRNA are chosen to induce proteins to be produced in bursts of size $b=k_2/k1_1$, i.e., $10$ protein molecules are synthesized from a single transcript on average.

\subsubsection{Method of moments and maximum entropy reconstruction}
%

We compute an approximation of the moments 
of species $M$ and $P$ 
up to order $4$ ($\maxentnom+1=4$) and $6$ ($\maxentnom+1=6$) 
using the MM as explained in Section~\ref{sec:mc}.  
The moments of $P$ are then used  to reconstruct the marginal probability distribution of $P$.
For instance, given 
the moments $\Ex{X_P^0}, \Ex{X_P^1}, \Ex{X_{P}^2}, \Ex{X_{P}^3}, \Ex{X_{P}^4}$
we approximate the distribution of
$P$ with $\tilde{\maxent}(x) \approx \distr \left( X_P=x \right)$ (see Section~\ref{s:maxent}).
Due to the high sensitivity of the maximum entropy method 
even to small approximation errors of the moments of highest order considered 
(in this case the approximation of $\Ex{X_{P}^4}$ and  $\Ex{X_{P}^6}$, respectively), 
we use moments only up to $\Ex{X_{P}^3}$ for the reconstruction.
The same holds for all results presented below.
In Figure~\ref{fig:bursty_protein_MC_reconstructions} we plot the two
reconstructed distributions.

\begin{figure*}
	\centering
		\subfloat{	
			\includegraphics[width=0.5\textwidth]
			{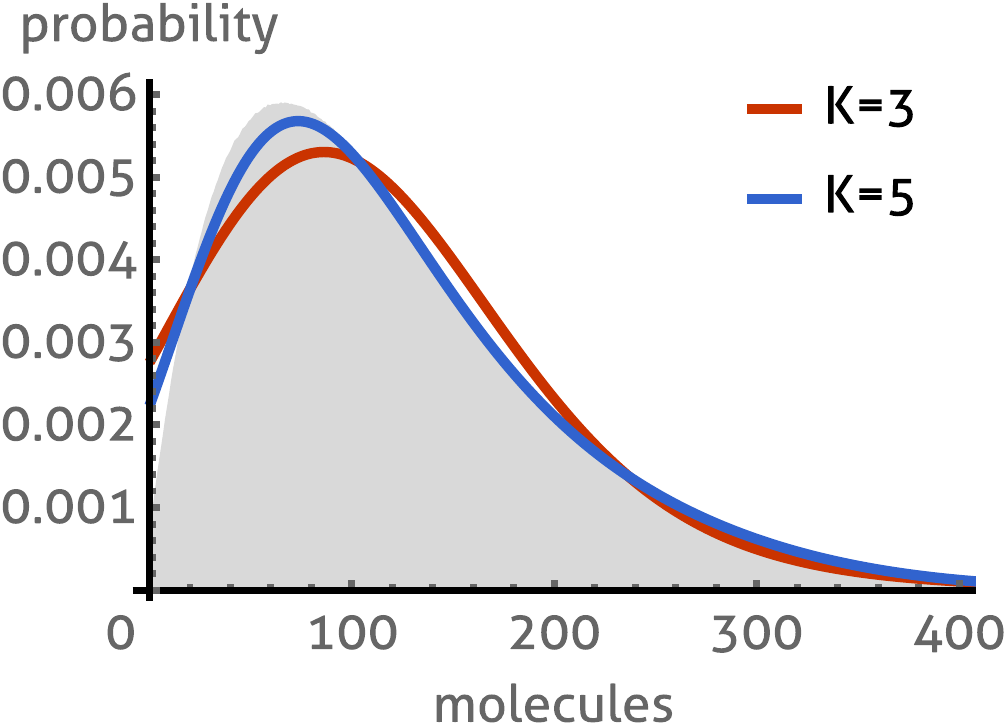}
			}
		\caption{
		\label{fig:bursty_protein_MC_reconstructions}
		\textbf{Bursty protein production:   reconstruction based on MM.} 
		The reconstructed   distribution (solid lines, cf. Eq.~\eqref{eq:maxent_reconstruction})
		is compared to the distribution estimated with stochastic simulations 
		(light gray), 
		where we use $\maxentnom = 3$  and $\maxentnom = 5$   for the moments
		in the maximum entropy method. We find that taking into account moments of higher order
		increases the accuracy significantly, in particular in the region of 0-200 proteins. 
		 }		
\end{figure*}

While for most parts of the support (including the tail) the distribution 
is accurately reconstructed even with 
$\maxentnom=3$,
the method is less precise when considering the probability of small copy numbers of $P$. To improve the reconstruction, 
we may use more moments, for instance $K=7$. 
However, in this case the moment equations become so stiff that the numerical integration fails completely. This happens due to a combination of  highly nonlinear derivatives of the rate function $f(\mol_P)$ with large values of the higher order moments.


\FloatBarrier

\subsubsection{Solution using the system size expansion}
\label{sec:enzSSE}

\begin{figure}
 \centering
 \includegraphics[width=1\textwidth]{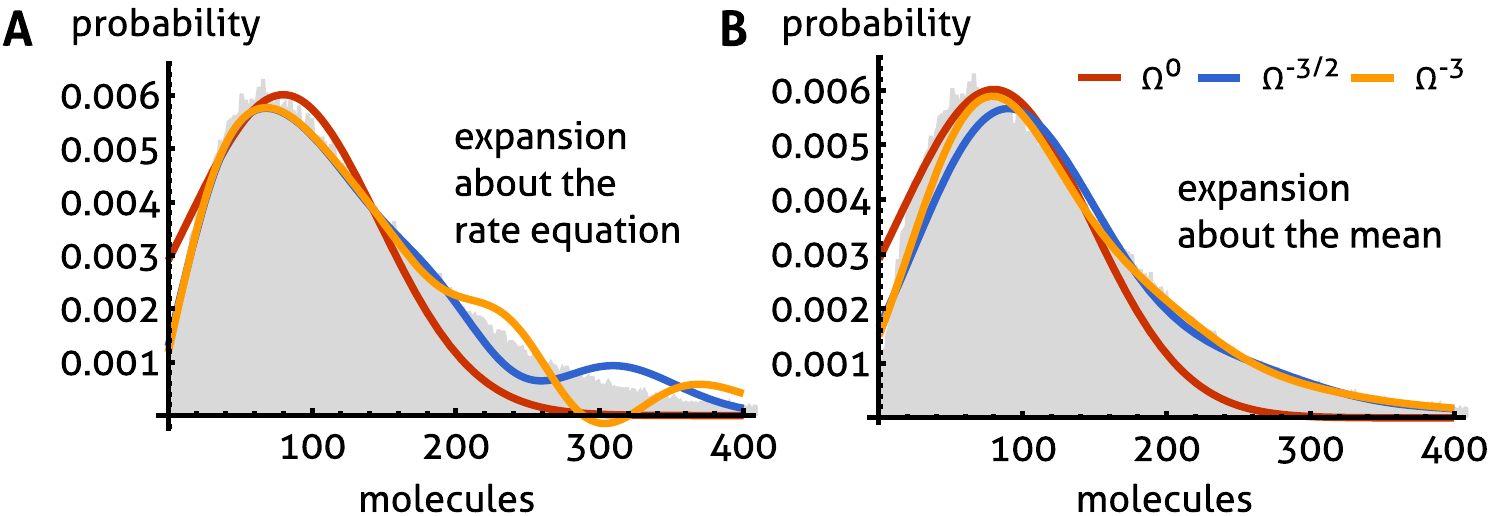}
 \caption{\textbf{Bursty protein production: system size expansion solution} (A) We compare the solution obtained using the system size expansion about the rate equation solution, Eq. (\ref{eqn:HermiteExpansion}), that is truncated after $\Omega^0$ (red), $\Omega^{-3/2}$ (blue) and $\Omega^{-3}$-terms (yellow line) to stochastic simulations (gray bars). We observe that the series yields large undulations and negative probabilities. (B) The system size expansion about the mean is shown when the series in Eq. (\ref{eqn:RenormalizedExpansion}) truncated after $\Omega^0$ (red), $\Omega^{-3/2}$ (blue) and $\Omega^{-3}$-terms (yellow line). We find that these approximations avoid undulations and converge rapidly with increasing truncation order to the distributions obtained from simulations using the SSA. 
 }
 \label{fig:bursty}
\end{figure}

The approximate solution to the distribution functions using the system size expansion as outlined in Section \ref{sec:sse} is available for networks of a single species only. We therefore concentrate on the case where the mRNA dynamics is much faster than the one of the protein, called the burst approximation \cite{shahrezaei2008}. It can be shown the reaction scheme (\ref{reactions:bursts}) follows the reduced CME
\begin{align}
 \label{appl:burstapprox}
 \frac{\mathrm{d}}{\mathrm{d}t} \distr(x) =&  \Omega  \sum_{z=0}^\infty (E_x^{-z}-1) h_0 \varphi(z) \distr(x) 
                    + \Omega (E_x^{+1}-1)  v_M \frac{x}{\Omega K_M+x} \distr(x),
\end{align}
where $E_x^{-z}$ is the step operator defined by $E_x^{-z}f(x)=f(x-z)$ for any function $f$. Note that protein synthesis occurs in random bursts $z$ following a geometric distribution $\varphi(z)= \frac{1}{1+b}\left(\frac{b}{1+b}\right)^z$ with average $b$. The relation between the parameters in scheme (\ref{reactions:bursts}) is  $h_0=k_0 k_2 / k_1$, $b=k_2/k_1$.
Within this description protein synthesis involves many reactions: one for each value of $z$ with probability $\varphi(z)$. The  coefficients in the expansion of the CME follow from Eq. (\ref{eqn:SSEcoeffs}), and are given by   
\begin{align}
 \label{appl:SSEcoeffs}
 \mathcal{D}_{n}^{m}= \delta_{m,0} h_0 \langle z^n \rangle_\varphi + (-1)^n v_M \frac{\partial^m}{\partial {[P]}^m}\frac{[P]}{K_M+[P]},
\end{align}
and $\mathcal{D}_{n,s}^{m}=0$ for $s>0$, where $[P]$ denotes the protein concentration according to the rate equation solution and $\langle z^n \rangle_\varphi=\sum_{z=0}^\infty z^n\varphi(z)=\frac{1}{1+b}\operatorname{Li}_{-n}(\frac{b}{1+b})$ denotes the average over the geometric distribution in terms of the polylogarithm function $\operatorname{Li}_{-n}(x) = \sum_{k=1}^\infty {k^n x^k}$. The rate equation (\ref{eqn:REs}) can be solved together with the linear noise approximation, Eq. (\ref{eqn:lyapunov}), in steady state conditions. For ${v_M}<b h_0$ the solution is
\begin{subequations}
\label{eqn:solEnzyme}
\begin{align}
 [P]=K_M \left(1-\frac{v_M}{b h_0}\right)^{-1}, \ \
 \sigma^2 = K_M (b+1) \varsigma  (\varsigma +1),
\end{align}
where we have defined by $\varsigma=\frac{[P]}{K_M}$ the reduced substrate concentration. In Fig. \ref{fig:bursty}A we show that the expansion performed about the solution the rate equation solution leads to large undulations, we therefore focus on the expansion about the mean. To this end, we have to take into account higher order corrections to the first two moments, we find $\langle \epsilon \rangle= \Omega^{-1/2}(1 + b)\varsigma$ and $\bar{\sigma}^2=\sigma^2+\Omega^{-1}{(b+1) \varsigma (b (\varsigma +2)+\varsigma +1)}$. The non-zero coefficients to order $\Omega^{-1}$ given by Eqs. (\ref{eqn:renormalizedCoeffSol}) then evaluate to
\begin{align}
 &\bar{a}_3^{(1)}=\frac{\sigma^2}{6} (2 b (\varsigma +1)+2 \varsigma +1), \notag\\
 &\bar{a}_4^{(2)}= \frac{\sigma^2}{4} \biggl[ (b+1)^2 \varsigma ^2+ (b+1) (2 b+1) \varsigma +\frac{1}{6} (6 b (b+1)+1) \biggr], \notag\\
 &\bar{a}_6^{(2)}= \frac{1}{2} \bigl[ \bar{a}_3^{(1)} \bigr]^2.
\end{align}
The coefficients to order $\Omega^{-3/2}$ can be obtained from Eqs. (\ref{eqn:RenSol}) and read
\begin{align}
 &\bar{a}_3^{(3)}=\frac{1}{6} (b+1) \varsigma  \left[2 (b+1)^2 \varsigma ^2+3 b (2 b+3) \varsigma +6 b (b+1)+3 \varsigma +1\right],\notag \\
 &\bar{a}_5^{(3)}=\frac{\bar{a}_3^{(1)}}{20} \left[
 1+12 b(b+1) +12 (b+1)^2 \varsigma ^2+12 (b+1) (2 b+1) \varsigma 
 \right],\notag\\
 &\bar{a}_7^{(3)}=\bar{a}_3^{(1)}\bar{a}_4^{(2)}, \ \
 \bar{a}_9^{(3)}=\frac{1}{6}\bigl[\bar{a}_3^{(1)}\bigr]^3.
\end{align}
\end{subequations}
The analytical form of these coefficients represents a particularly simple way of solving the CME. The approximation resulting from using these in Eq. (\ref{eqn:RenormalizedExpansion}) is shown Fig. \ref{fig:bursty}B (blue line). We find that these capture much better the true distribution obtained from exact stochastic simulation using the SSA (gray bars) than the linear noise approximation (red line). We find that including higher order terms in Eqs. (\ref{eqn:RenSol}) helped to improve the agreement. The resulting expressions turn out to be more elaborate and are hence omitted. This agreement is also confirmed quantitatively using the absolute and relative errors (see previous section for definition) given in the caption of Fig. \ref{fig:bursty}.

\FloatBarrier
\subsection{Example 2: Cooperative self-activation of gene expression}
\label{sec:selfrep}

As a second application of our methods we consider the regulatory dynamics of a single gene inducing its own, leaky expression. We therefore consider the case where gene activation occurs by binding of its own protein $P$ to two independent sites
\begin{align}
\label{reaction:sr1}
&G+P \xrightleftharpoons[k_{-1}]{k_1} G^\ast,\notag\\
&G^\ast+P \xrightleftharpoons[k_{-2}]{k_2} G^{\ast\ast}.
\end{align}
Here, $G$, $G^\ast$ and $G^{\ast\ast}$ denote the respective genetic states with increasing transcriptional activity leading to a cooperative form of activation which is modeled explicitly using mass-action kinetics. In effect, there are three gene states, corresponding to zero, one or two activators bound. Translation of a transcript denoted by $M$ therefore must occur via one of the following reactions
\begin{align}
\label{reaction:sr2}
&G \xrightarrow{k_G} G+M, \notag\\
&G^\ast \xrightarrow{k_{G^\ast}} G^\ast+M,\notag\\
&G^{\ast\ast} \xrightarrow{k_{G^{\ast\ast}}} G^{\ast\ast}+M,
\end{align}
where $k_G$ denotes the basal transcription rate, $k_{G^\ast}$ the transcription rate of the semi-induced state ${G^\ast}$, and $k_{G^{\ast\ast}}$ the rate of the fully induced gene. Finally, by the standard model of translation and neglecting active degradation we have
\begin{align}
\label{reaction:sr3}
&M \xrightarrow{k_3} M+P, \notag\\
&M\xrightarrow{k_4} \varnothing, \ \
P\xrightarrow{k_5} \varnothing,
\end{align}
In the following two parameter sets listed in Table~\ref{tab:PosFB} leading to moderate and low protein numbers are considered. As we shall see the protein distributions are multimodal in both cases representing ideal test cases for the distribution reconstruction using 
conditional moment closures and the conditional system size expansion.

\subsubsection{Method of conditional moments and maximum entropy reconstruction}
\begin{figure*}[t]
	\centering
	\includegraphics[width=\textwidth]{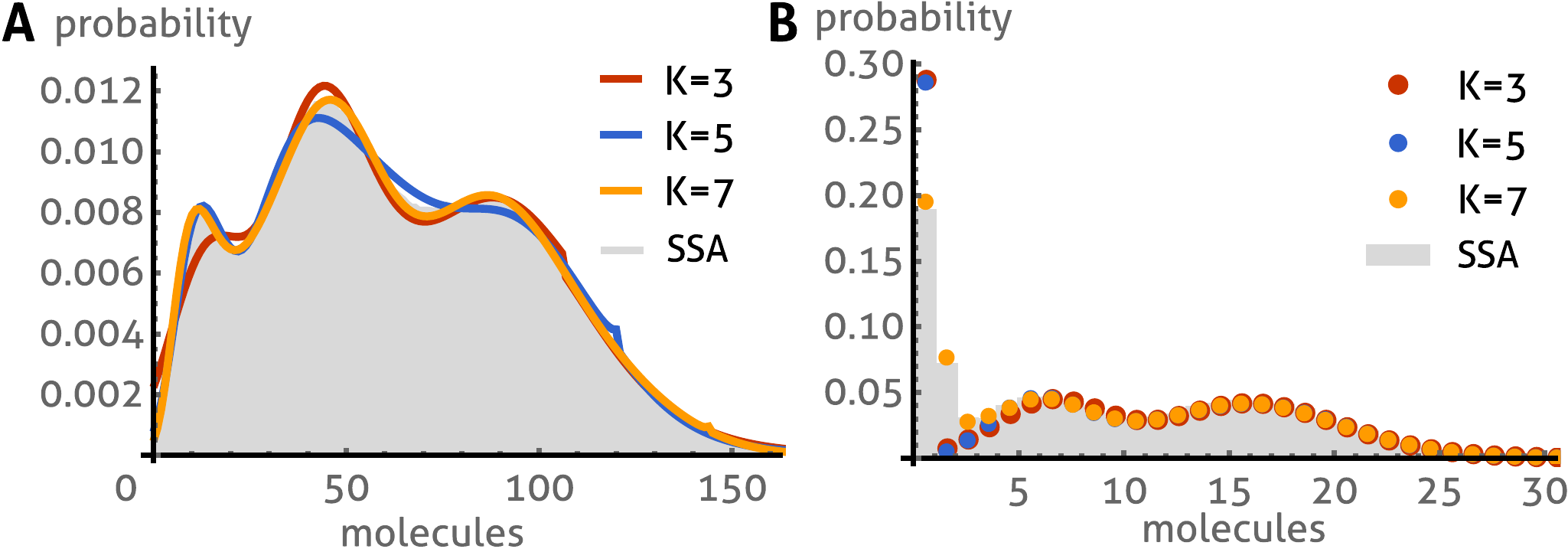}
		\caption{
		\label{fig:geneselfA_mcm_reconstructions}
		\textbf{Self-activating gene:  reconstruction based on MCM}. 
		The reconstructed distribution (solid lines, cf. Eq.~\eqref{eq:mcm_maxent_reconstruction})
		for the case of moderate (A) and slow (B) protein production
		is compared to the distribution estimated with stochastic simulations 
		(light gray),
		where we use 
		$\maxentnom = 3$, 
		$\maxentnom = 5$,
		and $\maxentnom = 7$ 
		for the conditional moments in the maximum entropy method. Again we find a significant
		improvement if moments of higher order are taken into account, in particular in those regions
		where the shape of the distribution is complex such as the region for 0-5 proteins in case of B.
		}
		
\end{figure*}

We compare the distribution reconstruction using an approximation of the first $3$, $5$ and $7$ moments of all the species obtained by the MCM (see Section \ref{sec:mc}).
As for the previous case study, the values of the moments of   $P$ are used 
to reconstruct the corresponding marginal distribution.
However, here we use the conditional moments 
$\Ex{X^k|G=1}$, $\Ex{X^k|G^\ast=1}$, $\Ex{X^k|G^{\ast\ast}=1}$
  instead.
We construct the function $\tilde{\maxent}(x)$
that approximates the distribution of
$P$ by applying the law of total probability 
\begin{equation}
\label{eq:mcm_maxent_reconstruction}
\tilde{\maxent}(x) = \sum_{S \in \{G, G^\ast, G^{\ast\ast} \} } P \left( S=1 \right) \tilde{\maxent}(x|S=1).
\end{equation}
The results of the approximation are plotted in Figure~\ref{fig:geneselfA_mcm_reconstructions}.
As we can see, the multimodality of the distribution is captured pretty well, 
and the quality of the reconstructed distribution is quite good in particular when using 7 moments. 


\begin{table}[t]
\begin{center}
\begin{tabular}{c|c|c|c|c|c|c|c|c|c|c|}
parameter & $k_1$ & $k_{-1}$ & $k_2$ & $k_{-2}$ & $k_G$ & $k_{G^\ast}$ & $k_{G^{\ast\ast}}$ &
$k_3$ & $k_4$ & $k_5$ \\
\hline
set (A) & $5\times 10^{-4}$ &  $3 \times 10^{-3}$ & $5 \times 10^{-4}$ & 
 $2.5 \times 10^{-2}$ & $4$ & $12$ & $24$ & $1200$ & $300$ & $1$ \\ 
set (B) & $5\times 10^{-4}$ & $2 \times 10^{-4}$ & $5\times 10^{-4}$ & $2 \times 10^{-3}$ & $4$ & $60$ & $160$ & $30$ & $300$ & $1$ \\
\end{tabular}
\end{center}
\caption{The two parameter sets used to study the multimodal expression of a self-activating gene described by the reactions (\ref{reaction:sr1}-\ref{reaction:sr3}). Set (A) leads moderate protein levels while set (B) yields low protein levels. Note that we have set $\Omega=1$.}
\label{tab:PosFB}
\end{table}


\FloatBarrier

\subsubsection{Conditional system size expansion }
\label{sec:SSE_selfactivation}

An alternative technique to approximate distributions for gene regulatory networks with multiple gene states has been given by \emph{Thomas et al.} \cite{thomas2014}. The method makes use of timescale separation by grouping reactions into reactions that change the gene state and reactions that affect only the protein distributions. Based on a conditional variant of the linear noise approximation it was shown that when the gene transitions are slow, protein distributions are well approximated by Gaussian mixture distributions. Implicit in this approach was, of course, that the protein numbers are sufficiently large to justify an application of an linear noise approximation. 
We will here extend this framework considering higher order terms in the system size expansion to account for low number of protein molecules. 

To this end, we describe by the vector $\vec{G}$ one of the three gene states $G=(1,0,0)$, $G^{\ast}=(0,1,0)$, and $G^{\ast\ast}=(0,0,1)$ and by $\mol$ the number of proteins. We will assume that gene transitions between these states evolve slowly on a timescale $1/\mu$. 
Rescaling time via $\tau= t/\mu$, the CME on the slow timescale reads 
\begin{align}
\frac{\mathrm{d} \distr\left(\vec{G},\mol,\tau\right)}{\mathrm{d} \tau}
  =& ~\mu \mathcal{W}_0 (\vec{G}) \distr\left(\vec{G},\mol,\tau\right)
     + \mathcal{W}_1 \distr\left(\vec{G},\mol,\tau\right),
\label{eqn:CMEgeneGamma}
\end{align}
where $\mathcal{W}_0 (\vec{G})$ describes the reactions (\ref{reaction:sr2}) and (\ref{reaction:sr3}) in the burst approximation
\begin{align}
 \label{eqn:W0}
\mathcal{W}_0 (\vec{G})=& 
   \sum_{z=0}^\infty (E_\mol^{-z}-1) \, \vec{k}\cdot \vec{G} \, \varphi(z) + (E_\mol^{+1}-1)  k_4 \mol,
\end{align}
where $\vec{k}=(k_{G},k_{G^\ast},k_{G^{\ast\ast}})$
and $\mathcal{W}_1$ denotes the transition matrix of the slow gene binding kinetics given by the reactions (\ref{reaction:sr1}). Note that as before $\varphi(z)$ is the geometric distribution with mean $b={k_3}/{k_4}$.
Using the definition of conditional probability, we can write $\distr(\vec{G},\mol,\tau) = \distr(\mol\,|\vec{G},\tau)\distr(\vec{G},\tau)$ which transforms Eq.~(\ref{eqn:CMEgeneGamma}) into
\begin{align}
\label{eqn:conditionalCME}
&\frac{\mathrm{d} {\distr}(\mol|\vec{G},\tau)}{\mathrm{d} \tau}\distr(\vec{G},\tau) +
\distr(\mol|\vec{G},\tau)\frac{\mathrm{d} \distr(\vec{G},\tau)}{\mathrm{d} \tau} \notag \\
&\qquad= \mu \, \distr(\vec{G},\tau) \,
 \mathcal{W}_0(\vec{G}) \distr(\mol|\vec{G},\tau)  + \mathcal{W}_1 \distr(\mol|\vec{G},\tau)\distr(\vec{G},\tau).
\end{align}
Marginalizing the above equation we find
\begin{align}
\frac{\mathrm{d} \distr(\vec{G},\tau)}{\mathrm{d} \tau} &= \left(\sum_{\mol=0}^\infty \mathcal{W}_1\, \distr(\mol|\vec{G})\right) \distr(\vec{G},\tau),
\end{align}
where the term in brackets is a conditional average of the slow dynamics over the protein concentrations. In steady state conditions the above is equal to the equations $0 = \distr(G) [P|G] - \distr(G^\ast) K_1$, $0 = \distr(G^\ast) [P|G^\ast]- K_2\distr(G^{\ast\ast})$
and $\distr(G^{\ast\ast})=1-\distr(G)-\distr(G^\ast)$ by conservation of probability. Here $[P|\vec{G}]$ denotes the conditional protein concentration that remains to be obtained from $\distr\left(\mol|\vec{G}\right)$. The steady state solution can be found analytically 
\begin{align}
 &\distr(G) = \frac{K_1 K_2}{K_1 K_2+K_2 [P|G]+[P|G] [P|G^\ast]},\notag\\
 &\distr(G^\ast) = \frac{K_2 [P|G]}{K_1 K_2+K_2 [P|G]+[P|G] [P|G^\ast]},\notag\\
 &\distr(G^{\ast\ast}) = \frac{[P|G] [P|G^\ast]}{K_1 K_2+K_2 [P|G]+[P|G] [P|G^\ast]},
\end{align}
where $K_1=\frac{k_{-1}}{k_1}$ and $K_2=\frac{k_{-2}}{k_2}$ are the respective association constants of the DNA-protein binding.
The protein distribution is then given by a weighted sum of the probability that a product is found given a particular gene state, times the probability of the gene being in that state:
\begin{align}
 \label{eqn:bayes_posterior}
 \distr(\mol) = \sum_{\vec{G}}\distr(\mol\,|\vec{G})\distr(\vec{G}).
\end{align}
It is however difficult to obtain $\distr(\mol\,|\vec{G})$ analytically, we will therefore employ the limit of slow gene transitions ($\mu\to\infty$) in Eq. (\ref{eqn:conditionalCME}) to obtain
\begin{align}
 \label{eqn:fastCME}
 \mathcal{W}_0(\vec{G}) \,\distrCond\left(\mol|\vec{G}\right) = 0,
\end{align}
where $\distrCond\left(\mol|\vec{G}\right)=\lim_{\mu\to\infty} \distr\left(\mol|\vec{G},\tau\right)$. We can now perform the system size expansion for the conditional distribution $\distrCond(\mol\,|\vec{G})$ that is determined by Eq. (\ref{eqn:fastCME}) using the ansatz 
\begin{align}
 \label{eqn:conditionalansatz}
  \left.\frac{\mol}{\Omega}\right|\vec{G} = [P|\vec{G}]+\Omega^{-1/2} \epsilon|\vec{G},
\end{align}
for the conditional random variables describing the protein fluctuations. The coefficients in the expansion of the CME (\ref{eqn:fastCME}) are
\begin{align}
 \label{eqn:SSEcoeffs_gene}
 \mathcal{D}_{n}^{m}= \delta_{m,0} \left( \vec{k}'\cdot \vec{G}\, \langle z^n \rangle_\varphi + (-1)^n k_1 [P|\vec{G}] \right) + \delta_{m,1} (-1)^n k_1 ,
\end{align}
with $\vec{k}'=\vec{k}/\Omega$ and $\mathcal{D}_{n,s}^{m}=0$ for $s>0$. The solution of the rate equation and the conditional linear noise approximation are given by
\begin{align}
 \label{eqn:LNAres}
 [P|\vec{G}]= \vec{k}'\cdot \vec{G} \,\frac{b}{k_5}, \ \ \sigma^2_{P|\vec{G}}= [P|\vec{G}](1 + b),
\end{align}
respectively.
Note that there are no further corrections in $\Omega$ to this conditional linear noise approximation because the conditional CME (\ref{eqn:fastCME}) depends only linearly on the protein variables. The conditional distribution can now be obtained using the result (\ref{eqn:HermiteExpansion}). Associating with $\pi_0(\epsilon\,|\vec{G})$ a centered Gaussian of variance as given in Eq. (\ref{eqn:LNAres}), we find to order $\Omega^{-1}$,
\begin{subequations}
 \label{eqn:cSSE}
\begin{align}
 \label{eqn:cSSE1}
 \distrCond(\epsilon|\vec{G}) = \pi_0(\epsilon&|\vec{G})\biggl[ 1 + \Omega^{-1/2} a_3^{(1)}(\vec{G}) \psi_{3,\vec{G}}(\epsilon) 
 \notag\\ &+ \Omega^{-1} a_4^{(2)}(\vec{G}) \psi_{4,\vec{G}}(\epsilon) + \Omega^{-1}a_6^{(2)}(\vec{G}) \psi_{6,\vec{G}}(\epsilon) \biggr] + O(\Omega^{-3/2}).
\end{align}
By the definition given before Eq. (\ref{eqn:HermiteExpansion}) the polynomials $\psi_{m,\vec{G}}(\epsilon)$ depend on the gene state via the conditional variance Eq. (\ref{eqn:LNAres}). The coefficients are obtained from Eqs. (\ref{eqn:renormalizedCoeffSol}) that lead to the particularly simple expressions
\begin{align}
 \label{eqn:cSSE2}
 &a_3^{(1)}(\vec{G})= \frac{1}{6} \left(2 b^2+3 b+1\right) [P|\vec{G}], \\
 &a_4^{(2)}(\vec{G})=\frac{1}{24} (b+1) \left(6 b^2+6 b+1\right) [P|\vec{G}], \ \
 &a_6^{(2)}(\vec{G})=\frac{1}{2} \bigl[ a_3^{(1)}(\vec{G})\bigr]^2.
\end{align}
\end{subequations}
Finally, we remark that $\Pi(\mol\,|\vec{G})$ and $\Pi(\epsilon\,|\vec{G})$ are related by $\Pi(\mol\,|\vec{G})=\Omega^{-1/2} \Pi(\epsilon=\Omega^{-1/2} \mol - \Omega^{1/2} [P|\vec{G}] \,|\vec{G})$. 

\begin{figure}[t]
 \centering
 \includegraphics[width=\textwidth]{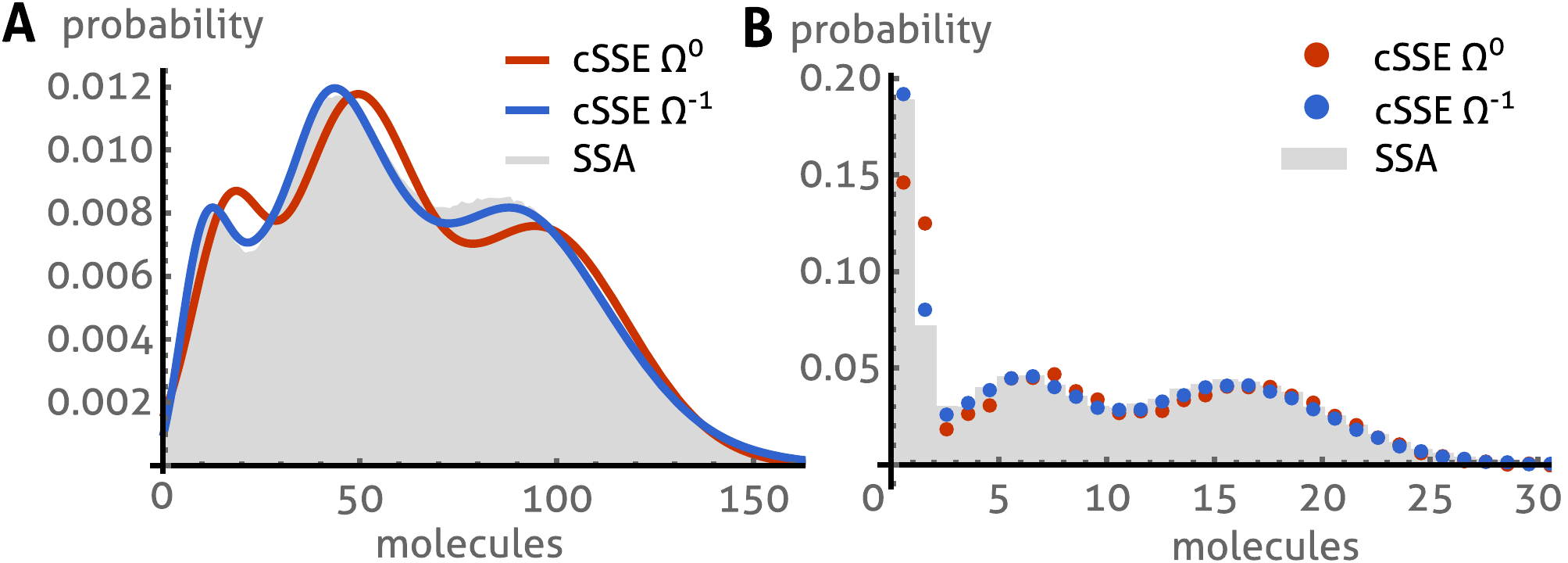}
 \caption{\textbf{Self-activating gene: conditional system size expansion.} The conditional system size expansion (cSSE), Eqs. (\ref{eqn:bayes_posterior}) and (\ref{eqn:cSSE}), for the moderate (A) and slow (B) protein production  is compared to stochastic simulations (SSA). While the conditional linear noise approximation ($\Omega^0$) captures the multimodal character of the distribution, it misses the precise location of its modes. We find that the $\Omega^{-1}$-estimate of the cSSE given by Eqs. (\ref{eqn:cSSE}) agrees much better with the SSA. 
 (B) The discrete approximation of Ref. \cite{thomas2015}, see main text for details, is shown when truncated after $\Omega^0$ and $\Omega^{-1}$ terms. The analytical form of the coefficients in Eq. (\ref{eqn:cSSE2}) has been used with parameter set B in Table \ref{tab:PosFB}. 
}
 \label{fig:cSSE}
\end{figure}

In Fig. \ref{fig:cSSE}A, this conditional system size expansion is compared to stochastic simulation using the SSA. We find that the conditional linear noise approximation, Eq. (\ref{eqn:cSSE}), truncated after $\Omega^0$ captures well the multimodal character of the distributions but misses to predict the precise location of its modes. In contrast, the conditional approximation of Eq. (\ref{eqn:cSSE}) taking into account up to $\Omega^{-1}$ terms accurately describes the location of these distribution maxima. We note however that a continuous approximation such as Eq. (\ref{eqn:cSSE1}) may fail in situations when the effective support of the conditional distributions represents the range of a few molecules. Such case is depicted in Fig. \ref{fig:cSSE}B. In this case the distributions are captured better by an approximation with discrete support as has been given in Ref. \cite{thomas2015}, Eqs. (35) and (36) therein. The resulting approximation using the analytical form of the coefficients (\ref{eqn:cSSE2}, blue dots) is in excellent agreement with stochastic simulation performed using the SSA (gray bars). These findings highlight clearly the need to go beyond the conditional Gaussian approximation for the two cases studied here.

\subsection{Comparison of numerical results}

\begin{figure}
 \includegraphics[width=\textwidth]{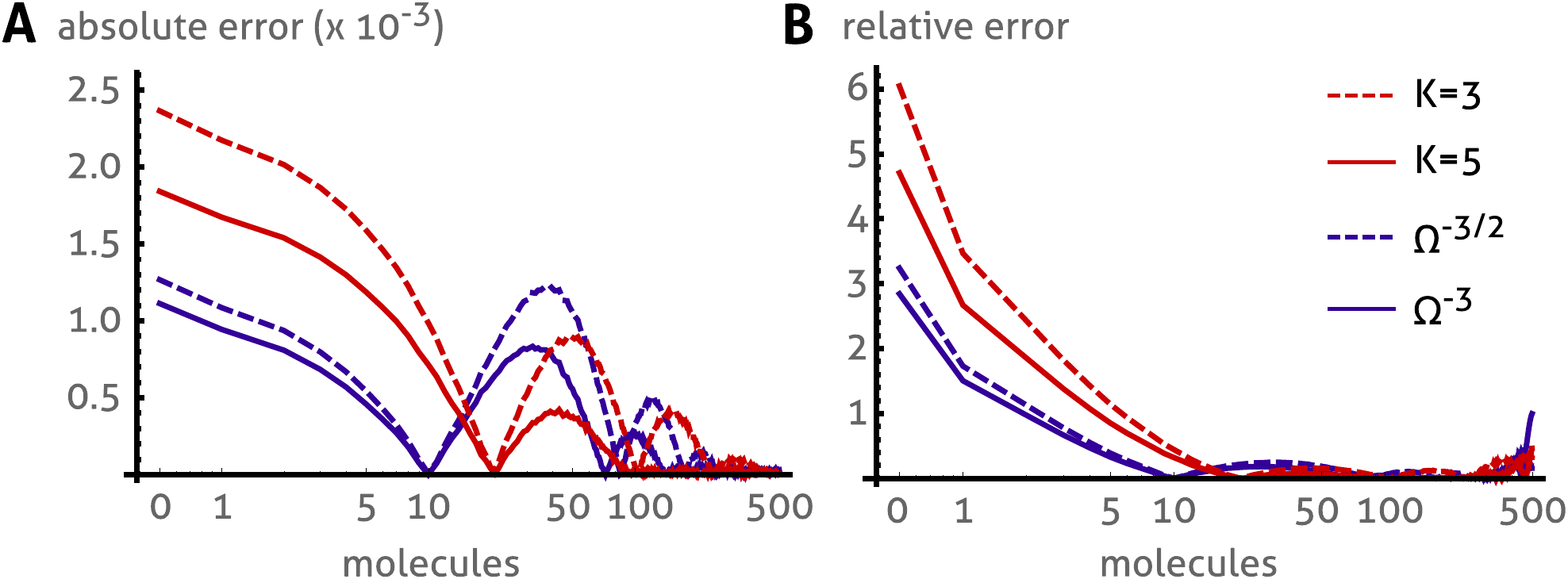}
 \caption{\textbf{Bursty protein production: absolute and relative error.} 
 We plot the absolute (A) and relative (B) errors of the MM (for  
 moments up to order $K=3$ and $K=5$, Fig. \ref{fig:bursty_protein_MC_reconstructions}) and the SSE (truncated after $\Omega^{-3/2}$ and $\Omega^{-3}$ terms, Fig. \ref{fig:bursty}).
 The SSE for this example yields the percentage error $|| \epsilon ||_V$ to be $5.1 \%$ ($\Omega^{-3/2}$) and $2.8 \%$ ($\Omega^{-3}$) while the moment based approach yields $5.6 \%$ ($K=3$) and $2.0\%$ ($K=5$).
 Both approaches become more accurate as more moments
 or higher order terms in the SSE are taken into account. For both methods, the maximum errors occurs at zero molecules.
 }
 \label{fig:err1}
\end{figure}

\begin{figure}
 \includegraphics[width=\textwidth]{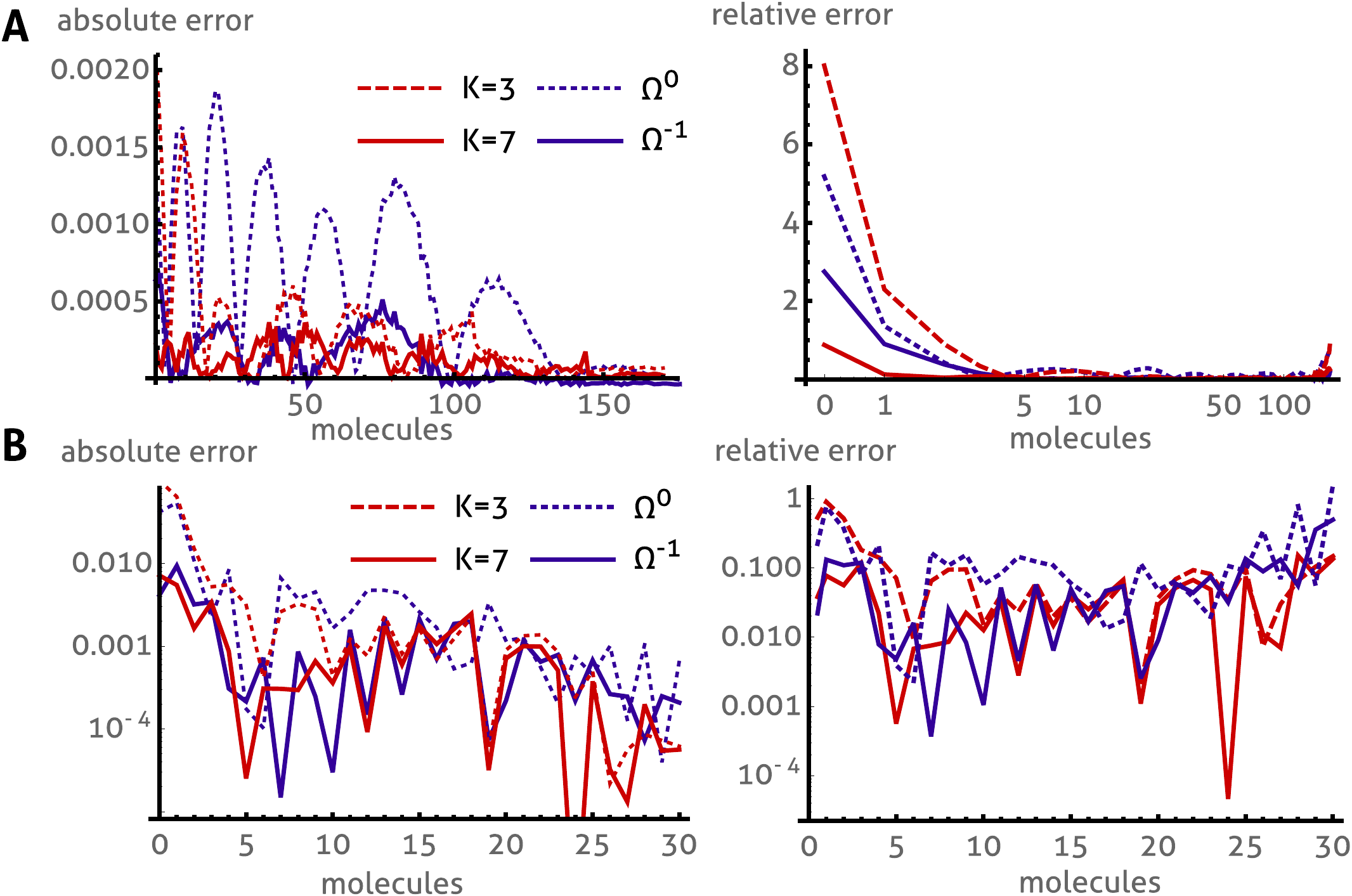}
 \caption{\textbf{Self-activating gene: absolute and relative error. }
 We plot the absolute (left) and relative (right) errors of the MCM (for  
 moments up to order $K=3$ and $K=7$) and the cSSE  (truncated after $\Omega^{0}$ and $\Omega^{-1}$ terms).
 For moderate protein production (A), corresponding to the distributions shown in Fig. \ref{fig:geneselfA_mcm_reconstructions}A and \ref{fig:cSSE}A, the MCM yields a percentage error $|| \epsilon ||_V$ of $2.4 \%$ ($K=3$) and $1.0 \%$ ($K=7$) while the cSSE yields  $5.4 \%$ ($\Omega^0$) and $1.4 \%$ ($\Omega^{-1}$), respectively.
  For slow protein production (B), corresponding to the distributions shown in Fig. \ref{fig:geneselfA_mcm_reconstructions}B and \ref{fig:cSSE}B, we find $|| \epsilon ||_V=10.9 \%$ ($K=3$) and $1.7 \%$ ($K=7$) for the MCM as well as $|| \epsilon ||_V=8.3 \%$ ($\Omega^0$) and $1.9 \%$ ($\Omega^{-1}$) for the cSSE, respectively.
  }
 \label{fig:err2}
\end{figure}

For the first case study, we calculated absolute $| \distr(x) - \distr^\ast(x) |$ and relative errors $| \distr(x) - \distr^\ast(x) |/\distr(x)$ between the exact distribution $\distr(x)$, which was estimated from simulations via the SSA,
and the distribution approximation obtained either via the MM or via the SSE denoted by $\distr^\ast(x)$. The results for the first case study are shown in Fig. \ref{fig:err1}. 
We observe that the maximum absolute and relative errors occur close to the boundary of zero molecules for both approximation methods. However, SSE is more accurate than MM here.
In this region a direct representation of the probabilities (as in the hybrid approaches) may be more appropriate.
 For measuring the overall agreement of the distributions we computed the percentage statistical distance 
\begin{align}
 || \epsilon ||_V = \frac{100\%}{2} \sum_{x=0}^\infty | \distr(x) - \distr^\ast(x) |.
\end{align}
This distance can also be interpreted as the maximum percentage difference between the probabilities of all possible events assigned by the two distributions \cite{levin2009,cover2012} and achieves its maximum ($100\%$ error) when the distributions do not overlap. The numerical values given in the caption of Fig. \ref{fig:err1} reveal that the estimation errors of the MM and the SSE decrease as more moments or higher order terms in the SSE are taken into account. 
The respective error estimates are of the same order of magnitude. However, the analytical solution obtained using the SSE, given in Section \ref{sec:enzSSE} including only low order terms is slightly more accurate than the MM with only few moments, while the MM with a larger number of moments becomes more accurate than the SSE including up to $\Omega^{-3}$-terms.

For the second case study, we find that the absolute and relative estimation errors of the method of moments and the SSE are of the same order of magnitude, compare Fig.~\ref{fig:err2}. 
However, we found that the method of moments including three conditional moments ($K=3$) is overall more accurate than the conditional linear noise approximation ($\Omega^0$).
The approximations become comparable as higher order conditional moments and higher orders in conditional SSE are employed. 
However, the method of moments including 7 conditional moments turned out to be slightly more accurate 
than the analytical SSE solution to order $\Omega^{-1}$ given in Section~\ref{sec:SSE_selfactivation}.

\FloatBarrier

\section{Discussion}
\label{sec:discussion}

We have here studied the accuracy of two recently proposed approximations for the probability distribution of the CME. The method of moments utilizes moment closure to approximate the first few moments of the CME from which the distribution is reconstructed via the maximum entropy principle. In contrast, the SSE method does not rely on a moment approximation but instead the probability distribution is obtained analytically as a series in a large parameter that corresponds roughly to the number of molecules involved in the reactions. Interestingly, our comparative study revealed that both methods yield comparable results. While generally both methods provide highly accurate approximations for the distribution tails and capture well the overall shape of the distributions, we found that for both methods the largest errors occur close to the boundary of zero molecules. We observed a similar behaviour when conditional moments or the conditional SSE were used. These discrepancies could be resolved by taking into account higher order moment closure schemes or, equivalently, by taking into account higher order terms in the expansion of the probability distribution.

From a computational point of view, the method based on moment closure is limited by the number of moments that can be numerically integrated due to the stiffness of the resulting high-dimensional ODE system. Our investigation showed that such difficulties are encountered when one closes the moment equations beyond the $8^{th}$ moment. In contrast, the analytical solution provided by the SSE technique does not suffer from these issues because it is provided in closed form. We note however that the SSE solution given here is limited to a single species while the method of moments has no such limitation. Moreover, the conditional SSE solution relies on timescale separation which the method of conditional moments does not assume.

The computational cost of the analytical approximation provided by the SSE is generally less than that of the moment based approach because it avoids numerical integration and optimization. This fact may be particularly advantageous when wide ranges of parameters are to be studied, as for instance in parameter estimation from experimental data. We note however that the moment based approach is still much faster than the one of the SSA because it avoids large amounts of ensemble averaging. Therefore the moment based approach may be preferable in situations where the SSE cannot be applied as we have mentioned above. We hence conclude that both approximation schemes provide complementary strategies for the analysis of stochastic behaviour of biological systems. Most importantly, they both provide much more computationally efficient strategies compared with simulation and numerical integration of the CME, preserving an high degree of accuracy, showing high potential for the analysis of large-scale models.


\printbibliography

\end{document}